# Computing extinction maps of star nulling interferometers

**François Hénault**

*Observatoire de la Côte d'Azur, Laboratoire Gemini, UMR 6203, CNRS*
*Avenue Nicolas Copernic, 06130 Grasse, France*
*Francois.Henault@obs-azur.fr*

*http://www.obs-azur.fr/*

**Abstract:** Herein is discussed the performance of spaceborne nulling interferometers searching for extra-solar planets, in terms of their extinction maps projected on-sky. In particular, it is shown that the designs of Spatial Filtering (SF) and Achromatic Phase Shifter (APS) subsystems, both required to achieve planet detection and characterization, can sensibly affect the nulling maps produced by a simple Bracewell interferometer. Analytical relationships involving cross correlation products are provided and numerical simulations are performed, demonstrating marked differences in the aspect of extinction maps and the values of attained fringes contrasts. It is concluded that depending on their basic principles and designs, FS and APS will result in variable capacities for serendipitous discoveries of planets orbiting around their parent star. The mathematical relationships presented in this paper are assumed to be general, i.e. they should apply to other types of multi-apertures nulling interferometers.

**OCIS codes:** (070.0700) Fourier optics and signal processing; (070.6110) Spatial filtering; (110.5100) Phase-array imaging systems; (350.1260) Astronomical optics

## References and links

1. M. Mayor and D. Queloz, "A Jupiter-Mass Companion to a Solar-Type Star," Nature **378**, 355-359 (1995).
2. M. Gillon, F. Courbin, P. Magain and B. Borguet, "On the potential of extrasolar transit planet survey," Astron. Astrophys. **442**, 731-744 (2005).
3. R.N Bracewell and R.H. MacPhie "Searching for non solar planets," Icarus **38**, 136-147 (1979).
4. A. Léger et al., "Could we search for primitive life on extrasolar planets in the near future ? The Darwin project," Icarus **123**, 249-255 (1996).
5. J.R.P. Angel and N.J.Woolf, "An imaging nulling interferometer to study extrasolar planets," Astrophys. J. **475**, 373-379 (1997).
6. O. Absil, R. den Hartog, P. Gondoin, P. Fabry, R. Wilhelm, P. Gitton and F. Puech, "Performance study of ground-based infrared Bracewell interferometers: Application to the detection of exozodiacal dust disks with GENIE," Astron. Astrophys. **448**, 787-800 (2006).
7. J.M. Le Duigou, M.Ollivier, A. Léger, F. Cassaing, B. Sorrente, B. Fleury, G. Rousset, O. Absil, D. Mourard, Y. Rabbia, L. Escarrat, F. Malbet, D. Rouan, R. Clédassou, M. Delpech, P. Duchon, B. Meyssignac, P.Y. Guidotti and N. Gorius, "Pegase: a space-based nulling interferometer," in Proc. of the SPIE **6265**, *Space Telescopes and Instrumentation I: Optical, Infrared, and Millimeter*, J.C. Mather, H.A. MacEwen and M.W.M. de Graauw eds., 62651M (2006).
8. M. Fridlund, "Darwin and TPF: technology and prospects," in Proc. of the SPIE **5491**, *New Frontiers in Stellar Interferometry*, W.A. Traub ed., 227-235 (2004).
9. *TPF-I Science Working Group Report*, JPL Publication 07-1, P.R. Lawson, O.P. Lay, K.J. Johnston and C.A. Beichman eds., Jet Propulsion Laboratory, California Institute of Technology, Pasadena, California (2007).
10. O.P. Lay, "Imaging properties of rotating nulling interferometers," Appl. Opt. **44**, 5859-5871 (2005).
11. B.F. Lane, M.W. Muterspaugh and M. Shao, "Calibrating an interferometric null," Astrophys. J. **648**, 1276-1284 (2006).
12. M. Ollivier and J.M. Mariotti "Improvement in the rejection rate of a nulling interferometer by spatial filtering," Appl. Opt. **36**, 5340-5346 (1997).
13. S. Shaklan and F. Roddier, "Coupling starlight into single-mode fiber optics," Appl. Opt. **27**, 2334-2338 (1988).

14. C. Ruilier and F. Cassaing, "Coupling of large telescope and single-mode waveguides," J. Opt. Soc. Am. A **18**, 143-149 (2001).
15. Y. Rabbia, J. Gay, J.P. Rivet and J.L. Schneider, "Review of Concepts and Constraints for Achromatic Phase Shifters," in *Proceedings of GENIE-DARWIN Workshop - Hunting for Planets*, H. Lacoste ed., ESA SP-522 (European Space Agency, 2003), 7.1.
16. F. Hénault, "Design of achromatic phase shifters for spaceborne nulling interferometry," Opt. Lett. **31**, 3635-3637 (2006).
17. V. Coudé Du Foresto, G. Perrin, C. Ruilier, B.P. Mennesson, W.A. Traub and M.G. Lacasse, "FLUOR fibered instrument at the IOTA interferometer," in Proc. of the SPIE **3350**, *Astronomical Interferometry*, R.D. Reasenberg ed., 856-863 (1998).
18. B. Mennesson, M. Ollivier and C. Ruilier, "Use of single-mode waveguides to correct the optical defects of a nulling interferometer," J. Opt. Soc. Am. A **19**, 596-602 (2002).

## 1. Introduction

Since the first discovery in 1995 of an extra-solar planet using the radial velocity method by Mayor [1], the number of similar recognized space bodies progressed outstandingly, now exceeding the emblematic number of 200. In parallel, the development of transit detection techniques, either ground or space borne [2], sounds as the promise of new spectacular breakthrough for the total number and types of detected planets. However, the astronomical challenge of our century will undoubtedly be the direct observation of telluric, Earth-like planets suitable for harbouring extra-terrestrial life. This stimulating goal implies that a spectral analysis of the exo-planet atmosphere can be made available. For that purpose, a very promising technique seems to be a "nulling" interferometer working in the infrared spectral range (where the contrast between the planet and its parent star is more favourable than in the visible band), as originally proposed by Bracewell and MacPhie [3].

Bracewell's interferometer consists of a pair of telescopes – denoted $T_1$ and $T_2$ in Fig. 1 – rotating around an axis $O_SZ$, located at mid-distance between $T_1$ and $T_2$ and directed at the star's center. Both telescopes are creating a constructive and destructive fringe pattern projected on the sky, herein called "extinction map". Exo-planets become detectable when their parent star is deeply occulted by the dark central fringe, while luminous flux emitted or reflected by the planet is modulated as both telescopes are rotated around $O_SZ$-axis. The optical beams then enter the recombination unit, which has to fulfil two main functions that are first, to add an achromatic phase shift equal to $\pi$ along one interferometer arm (e.g. $T_2$ arm in Fig. 1), and second to combine both telescope beams. After a few decades, several alternative concepts of nulling interferometers have been studied and proposed by various authors [4-7], and space administrations like European Space Agency (ESA) or NASA finally selected two major projects, respectively DARWIN and TPF-I (Terrestrial Planet Finder Interferometer) [8]. Summarizing all lessons learned from these studies, either theoretical or experimental, is clearly not the scope of this manuscript, and the reader is invited to refer for instance to the TPF-I Science Working Group Report [9]. Here below is dealt with a particular topic, which concerns the relationship between on-sky extinction maps actually generated by a nulling interferometer, and the practical designs of some of its essential subsystems, namely the Spatial Filtering (SF) and Achromatic Phase Shifter (APS) devices. The issue is summarized as follows.

- Computing rejection ratios of a simple Bracewell interferometer is straightforward when punctual detectors are considered, which is the case in a vast majority of studies. However, when more subtle questions are arising, such as the calibration and phase chopping of double crossed Bracewell nullers [10,11], the estimation of realistic nulling maps taking into account actual detector size becomes an important matter.

- Similarly, in order to loosen some drastic image quality requirements that would be cast upon opto-mechanical designs, nulling interferometers are equipped with spatial filtering devices, that are either simple pinholes located at the interferometer focus [12], or single mode optical fibers [13,14]. It is likely that those SF devices will alter the expected extinction maps.

- Finally, nulling interferometers also incorporate one or several APS, whose function is to add along one or several arms a constant phase-shift of half-period over the whole spectral range of interest. The reader may refer to [15] in order to get a complete and synthetic view of possible APS designs. In fact, it has already been mentioned that the concept of selected APS also affects the generated extinction maps [16]. Here can be distinguished two different APS families:

  1) APS which do not change geometrical characteristics of the optical beams such as angular Field of View (FoV) directions and pupil axes.

  2) APS which modify that beam geometry, with the consequence that FoVs are being inverted with respect to the main optical axis (thus creating a couple of images for one single planet in Bracewell configuration), and pupils are flipped with respect to the central axis. We designate this family as "FoV-reversal" or "Pupil-flip" type.

In section 2 are developed some theoretical formulae of extinction maps produced by a single Bracewell interferometer, equipped with different types of FS and APS devices. Numerical simulation results are provided and commented in section 3, while a short conclusion is given in section 4.

## 2. Theory

For a Bracewell interferometer incorporating no FoV-reversal APS, the on-sky extinction map $E(u,v)$ is often estimated assuming punctual detectors, and neglecting diffraction effects arising from the finite diameter D common to both telescopes $T_1$ and $T_2$, as depicted in Fig. 1. Let $\lambda$ be the wavelength of the diffracted beams (supposed to be monochromatic) and B the interferometer baseline. A very simple relationship is soon retrieved:

$$E(u,v) = A_0^2 \sin^2(\phi) = A_0^2 \sin^2(\pi Bu/\lambda) \qquad (1)$$

if $A_0$ is the amplitude of light on both $T_1$ and $T_2$ entrance pupils, and $\phi$ is standing for the optical path difference $\pi Bu/\lambda$. However when detector size, diffraction generated by the telescopes and the natures of FS and APS subsystems are to be considered, it is more convenient to transpose $E(u,v)$ in the detection plane of the interferometer O'X'Y' (see Fig. 1). Actually, for each direction in the sky of angular coordinates $(u,v)$ corresponds a point of coordinates $(x' = F u, y' = F v)$ in the detection plane, where F is the focal length of the interferometer. Hence the complex amplitude generated in the image plane writes:

$$A(u,v,x',y') = A_0[\exp(-i\phi) - \exp(i\phi)] \hat{B}_D(x'-Fu, y'-Fv)/2 \qquad (2)$$

where $\hat{B}_D(x',y')$ is the Fourier transform of the "pillbox" function $B_D(x,y)$ of diameter D, which is uniformly equal to 1 inside a circle of radius D/2, and zero outside of this area. Its Fourier transform is analytically defined as:

$$\hat{B}_D(x',y') = 2 J_1(\pi D\rho/\lambda F) / (\pi D\rho/\lambda F) \qquad (3)$$

with $\rho^2 = x'^2 + y'^2$ and $J_1$ is the type-J Bessel function at the first order – it must be noticed that $\hat{B}_D^2(x',y')$ is nothing else than the Point Spread Function (PSF) generated by one given, individual telescope aperture, which is the well-known Airy spot. This remains true as long as each aperture exhibits no central obscuration, and the latter assumption remains applicable in the whole following sections including numerical results. In the presence of central obscuration, Eq. (3) is no longer valid and $\hat{B}_D^2(x',y')$ stands for the actual telescope PSF.

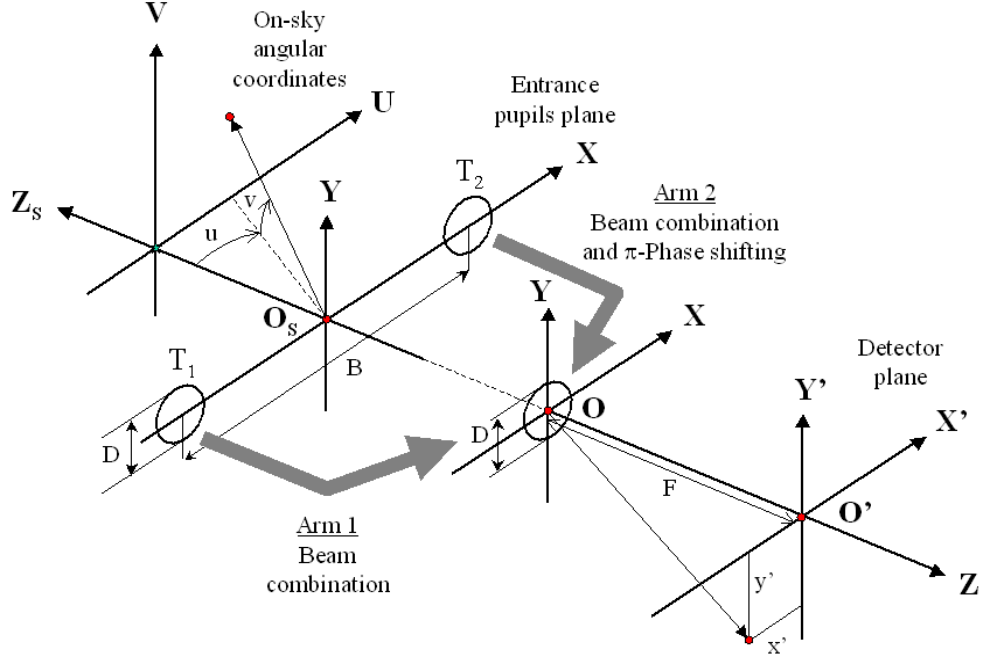

Fig. 1. Computing sky maps of a Bracewell interferometer

Eq. (2) will be the basis for evaluating extinction maps of the Bracewell interferometer, whatever the types of used SF and APS devices. We shall first consider the case when spatial filtering is achieved by means of a pinhole, and detector sensing area lies just behind it. In sections 2.3 and 2.4 will be studied effects of Single-Mode Fiber (SMF) filtering.

*2.1 Pinhole filtering without Pupil-flip*

This is the simplest case, where optical power can be deduced from amplitude distribution $A(u,v,x',y')$ by multiplying it with its complex conjugate, and then integrating over the whole pinhole area. Let P denote the diameter of the pinhole, which is defined by the pillbox function $B_P(x',y')$. Hence an expression of the collected energy $E(u,v)$ is derived:

$$E(u,v) = A_0^2 \sin^2(\phi) \iint_{x',y'} |\hat{B}_D(x'-Fu,y'-Fv)|^2 B_P(x',y')\, dx'dy' \Big/ A_0^2 \iint_{x',y'} B_P(x',y')\, dx'dy' \qquad (4)$$

where the extinction map is normalized with respect to the geometrical surface of the pinhole. Let us rewrite Eq. (4) recognizing its denominator as a cross correlation represented by the symbol $\otimes$:

$$E(u,v) = \sin^2(\pi Bu/\lambda)\left[|\hat{B}_D|^2 \otimes B_P\right](Fu, Fv)\Big/ S_P \qquad (5)$$

with $S_P$ the circular pinhole area. We thus demonstrated that the extinction map of a Bracewell interferometer equipped with pinhole filters is proportional to the product of the fringe pattern of Eq. (1) with an envelope function $|\hat{B}_D|^2 \otimes B_P$, itself being the cross correlation product of the PSF of an individual telescope with the pinhole function, both projected on-sky. This has the practical consequence of reducing the dimensions of the useful interferometric FoV.

*2.2 Pinhole filtering associated to a Pupil-flip*

Let us now consider the case of Pupil-flip APS. For one given angular direction (u,v) those devices have the property to form twin symmetrical images of equal power in the interferometer image plane [15,16], respectively centred on the (Fu,Fv) and (-Fu,-Fv) coordinates. The diffracted amplitude A(u,v,x',y') is then equal to:

$$A(u,v,x',y') = A_0 \left[\exp(-i\phi)\hat{B}_D(x'\text{-}Fu,y'\text{-}Fv) - \exp(i\phi)\hat{B}_D(x'\text{+}Fu,y'\text{+}Fv)\right]/2 \qquad (6)$$

Obviously the overlap area between both images decreases as the sky-object is located far from the central star, yielding potential fringe contrast loss. Total energy collected by the pinhole and detector is then equal to the integral of the right-hand term in Eq. (6), multiplied with its complex conjugate and with the pinhole function $B_P$. Hence the following expression of E(u,v) can finally be established:

$$E(u,v) = \left\{ \left[\left|\hat{B}_D\right|^2 \otimes B_P\right](Fu,Fv) + \left[\left|\hat{B}_D\right|^2 \otimes B_P\right](-Fu,-Fv) \right.$$
$$\left. - 2\cos(2\phi)\iint_{x',y'} \hat{B}_D(x'\text{-}Fu,y'\text{-}Fv)\hat{B}_D(x'\text{+}Fu,y'\text{+}Fv)B_P(x',y')\,dx'dy' \right\} \Big/ 4S_P \qquad (7)$$

No extra analytical simplification seems feasible, and one has to compute the integral in Eq. (7) numerically, for each considered couple of angular coordinates (u,v) as will be done in section 3. Let us now pay attention to the case when spatial filtering is achieved by means of Single-Mode Fibers (SMF), which is a very popular technique in the field of nulling interferometry – though initially pushed forward by ground-based stellar interferometry.

*2.3 SMF filtering without Pupil-flip*

Filtering of wavefront errors affecting both interferometer arms using SMFs was suggested almost two decades ago [14], and later successfully demonstrated on the FLUOR ground instrument [17]. It is currently envisaged as the best candidate technology for SF devices in nulling interferometry [18]. SMF have the unique property of filtering beams amplitudes rather than their intensities, thus transforming image quality defects into losses of coupled power into the fibers, which can be balanced more easily between both interferometer arms. According to the theory [14] and using Gaussian approximation (implicitly meaning that the operating wavelength $\lambda$ is close to the cut-off wavelength $\lambda_C$), the coupled amplitude $A_1(u,v,x',y')$ from arm n°1 writes into the SMF entrance plane:

$$A_1(u,v,x',y') = A_0^2 \exp[-i\phi] \iint_{x',y'} \hat{B}_D(x'\text{-}Fu,y'\text{-}Fv)\exp\left[-(x'^2+y'^2)/R_F^2\right]dx'dy' \Big/ 2A_0^2 C \qquad (8)$$

where $R_F$ is the radius at $1/e^2$ – in intensity – of the SMF entrance lobe, and C is a normalization constant equal to:

$$C = \left[\iint_{x',y'}\left|\hat{B}_D(x'\text{-}Fu,y'\text{-}Fv)\right|^2 dx'dy'\right]^{1/2} \left[\iint_{x',y'}\exp\left[-2(x'^2+y'^2)/R_F^2\right]dx'dy'\right]^{1/2} = 2^{1/2}\lambda FR_F/D \qquad (9)$$

Again, we recognize Eq. (8) as a cross correlation, here operated between the telescope diffracted amplitude and SMF entrance lobe functions:

$$A_1(u,v,x',y') = \exp(-i\phi)\left\{\hat{B}_D \otimes \exp\left[-(x'^2+y'^2)/R_F^2\right]\right\}(Fu,Fv)/2C \qquad (10)$$

The expression of amplitude $A_2(u,v,x',y')$ coupled into arm n°2 is very similar to the previous relationship, simply replacing quantity exp[-i$\phi$] with -exp[i$\phi$]. Summing both complex amplitudes $A_1$ and $A_2$ and multiplying the result with its complex conjugate finally defines the total power effectively measured by the instrument at FoV angles (u,v). Then an expression of the extinction map E(u,v) becomes:

$$E(u,v) = \sin^2(\pi Bu/\lambda) \left| \hat{B}_D \otimes \exp\left[-(x'^2+y'^2)/R_F^2\right] \right|^2 (Fu, Fv) / C^2 \quad (11)$$

One can see that the obtained formula is quite similar to the pinhole filter case, showing the same fringe pattern associated to a modified envelope, while the basic SMF parameter $R_F$ appears on the cross correlation. Numerical computations of section 3 will confirm that the resulting extinction maps E(u,v) somewhat differ between pinholes and SMFs.

*2.4 SMF filtering with a Pupil-flip*

Combining SMF filtering with one or several APS exhibiting pupil-flip (or FoV-reversal) properties is probably the most complicated case, although not uncommon. Moreover the mathematical relationships provided in this section highlight an interesting property in the case of a two-telescopes, Bracewell interferometer. Here we first replace the pillbox function $B_D(x,y)$ defining the contours of the telescopes by a more general function denoted T(x,y), whose Fourier transform will be $\hat{T}(x',y')$ in the interferometer focal plane, and normalization factor $C_T$. Amplitudes diffracted along each interferometer arm and integrated over the whole SMF area are then equal to:

<u>Arm 1</u>  $A_1(u,v,x',y') = \exp[-i\phi] \iint\limits_{x',y'} \hat{T}(x'-Fu,y'-Fv) \exp\left[-(x'^2+y'^2)/R_F^2\right] dx'dy' \Big/ 2C_T$ (12a)

<u>Arm 2</u>  $A_2(u,v,x',y') = -\exp[i\phi] \iint\limits_{x',y'} \hat{T}(x'+Fu,y'+Fv) \exp\left[-(x'^2+y'^2)/R_F^2\right] dx'dy' \Big/ 2C_T$ (12b)

if we assume that a pupil-flip APS is incorporated to arm n°2. Summing both amplitudes and multiplying by the complex conjugate finally leads to the total energy collected by the SMF:

$$E(u,v) = \Big\{ \left| \hat{T} \otimes \exp\left[-(x'^2+y'^2)/R_F^2\right] \right|^2 (Fu, Fv) + \left| \hat{T} \otimes \exp\left[-(x'^2+y'^2)/R_F^2\right] \right|^2 (-Fu, -Fv)$$
$$- 2\cos(2\phi) \times \hat{T} \otimes \exp\left[-(x'^2+y'^2)/R_F^2\right](Fu, Fv) \times \hat{T} \otimes \exp\left[-(x'^2+y'^2)/R_F^2\right](-Fu,-Fv) \Big\} / 4C_T^2 \quad (13)$$

Eq. (13) represents the most general mathematical relation expressing the extinction map of a Bracewell interferometer equipped with SMF and FoV-reversal APS. It must be emphasized that if, as is generally the case, T(x,y) is an axis-symmetric function – i.e. T(-x,-y) = T(x,y) – relation (13) promptly reduces to Eq. (11). Hence we demonstrated that the Bracewell interferometer is not sensitive to the actual design of the APS, provided that single-mode optical fibers are used, pupil function is centro-symmetric, and a perfect alignment is achieved between SMF and pupils.

### 3. Numerical results and discussion

Eq. (5), (7) and (11) presented in previous sections provide a strong basis for computing rejection maps generated by a two-arms, Bracewell-type interferometer, since they are all well-suited to numerical computation, and two of them involve simple cross correlation products. Depending on the actual concepts of SF and APS components, we shall consider the three following configurations.

- <u>Case n°1</u>: Spatial filtering is achieved by means of a circular pinhole and the achromatic phase-shifter does not modify the axes of the input beams (i.e. no FoV-reversal or Pupil-flip is present).

- <u>Case n°2</u>: Same type of spatial filtering, but using an APS that introduces FoV-reversal.

- Case n°3: Spatial filtering is realized with a single-mode optical fiber. Then the real nature of the APS has no consequences on the interferometer extinction maps (we assume that the telescope pupils are axis-symmetric as in section 2.4).

For Cases n°1 and 2, it is assumed that the pinhole diameter P is matched to the first dark ring of the Airy spot of an individual telescope, i.e. $P = 2.44 \lambda F/D$. For the SMF case, the fiber core radius $R_F$ is adjusted to reach the maximal possible coupling efficiency at the FoV centre as described in Ref. [18]. Assuming telescope diameters D = 20 m, baseline B = 100 m and wavelength $\lambda$ = 1 µm, surface plots of three typical on-sky extinction maps E(u,v) are displayed on Fig. 2 to 4, respectively for Cases n°1, 2 and 3. For the sake of understanding, we also computed extinction maps for an extremely short baseline B = 50 m, and reproduced slices of E(u,v) along the U-axis in Fig. 5. All other input parameters remain unchanged.

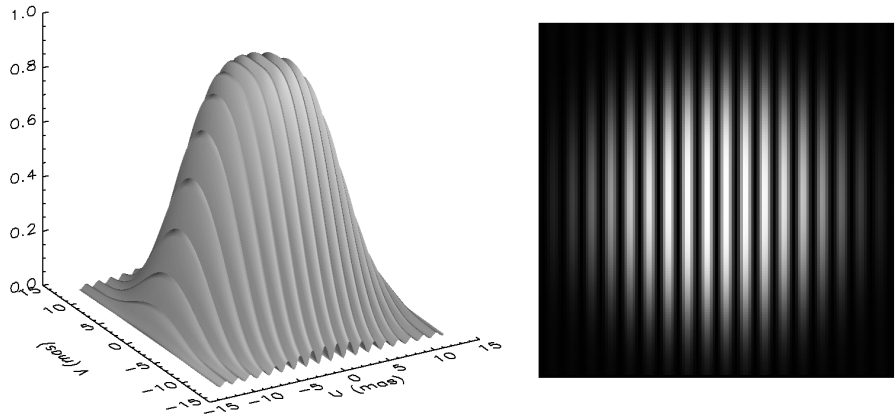

Fig. 2. Extinction map generated by pinhole filtering with no FoV reversal (left, 3D view; right, gray scale levels)

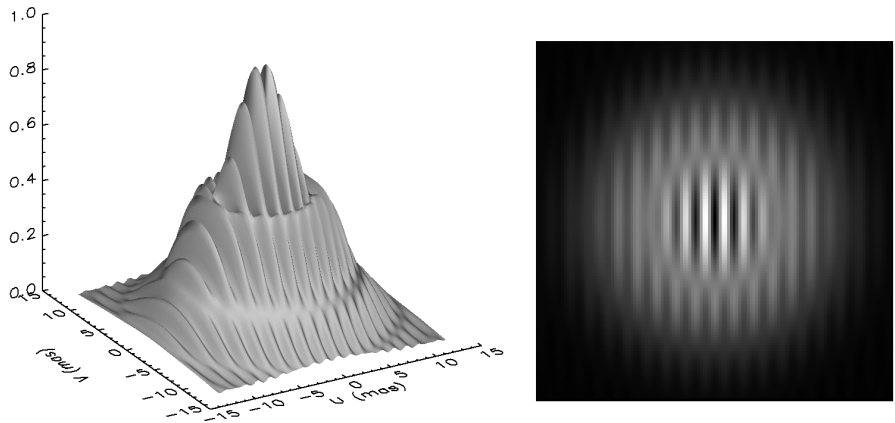

Fig. 3. Extinction map generated by pinhole filtering with FoV-reversal APS (left, 3D view; right, gray scale levels)

A careful analysis of numerical results illustrated in Fig. 2-3 and Fig. 5 partly confirms the conclusions of Ref. [16], also bringing to light one hidden error in the latter manuscript: due to a wrong normalization factor, all numerical values of rejection ratios produced by a pinhole-filtered, FoV-inverting interferometer have to be replaced with those computed and presented in the current paper. Therefore intensities differences between dark fringes and their neighbouring bright fringes, provided in Table 1 and plotted on Fig. 5, are lower than expected, and Signal-to-Noise Ratio (SNR) differences between both types of APS sensibly decrease. However this is probably not sufficient in order to rehabilitate the FoV-reversal APS, which still suffers from two major drawbacks, i.e.

- Only the central fringe can be used for star nulling, since rejection ratios of other dark fringes are not fully nulled (as is clearly seen on Fig. 5).

- The useful interferometer FoV (where fringes modulation attains its highest values) is globally shrunk by a factor around two.

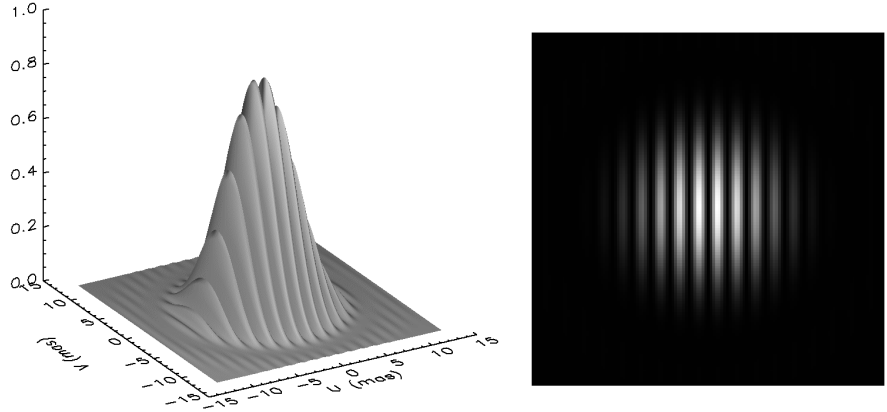

Fig. 4. Extinction map generated by SMF filtering (left, 3D view; right, gray scale levels)

Table 1. Variation of extinction ratio between dark and white fringes

| Spatial Filtering | FoV-reversal | Baseline (m) | Dark fringe number | | | | |
|---|---|---|---|---|---|---|---|
| | | | 0 (central) | 1 | 2 | 3 | 4 |
| Pinhole | No | 100 | 84 % | 84 % | 81 % | 75 % | 64 % |
| Pinhole | Yes | 100 | 82 % | 65 %* | 33 %* | 3 %* | 5 %* |
| SMF | Any type | 100 | 73 % | 63 % | 46 % | 28 % | 13 % |
| Pinhole | No | 50 | 84 % | 78 % | 56 % | 26 % | 7 % |
| Pinhole | Yes | 50 | 78 % | 24 %* | 3 %* | 4 %* | 0 %* |
| SMF | Any type | 50 | 72 % | 38 % | 9 % | 0 % | 0 % |

* Dark fringe not totally nulled.

Both previous disadvantages logically conduct to rule out technical designs or solutions based on pinhole filtering when associated to Pupil-flip APS, at least in planets serendipitous detection mode. Comparing Cases n°1 and 3 – respectively pinhole filtering with no FoV inversion and SMF – is also rich of information. Table 1 and both Fig. 3 and Fig. 5 show that SMF filtering is less efficient, in the sense that bright fringes, where the searched planets should ideally be located, present lower contrast, on one hand, and that the tendency visibly increases with the u angular coordinate, on the other hand. The fact is particularly noticeable for short interferometer baselines such as B = 50 m in Fig. 5 – although the case is most probably unrealistic, its main purpose is to focus the attention of the reader on both previous properties. From a mathematical point of view, they may originate from Eq. (11), where the cross correlation product is realized before extraction of its square modulus. We then retrieve at a lesser scale the second drawback of Case n°2, which consists in a significant reduction of the useful FoV of the nulling interferometer. As in Ref. [16], the major consequence may be that serendipitous discoveries are more difficult to achieve while using SMF filtering rather than simple pinholes. Hence the latter would be better suited to blind planet searching, while SMF might be kept for characterisation of planets whose existence has already been proved by other techniques. Reopening a complete trade-off between pinholes and SMF for nulling interferometry is however not the scope of this paper that mainly aims at providing additional elements to the discussion.

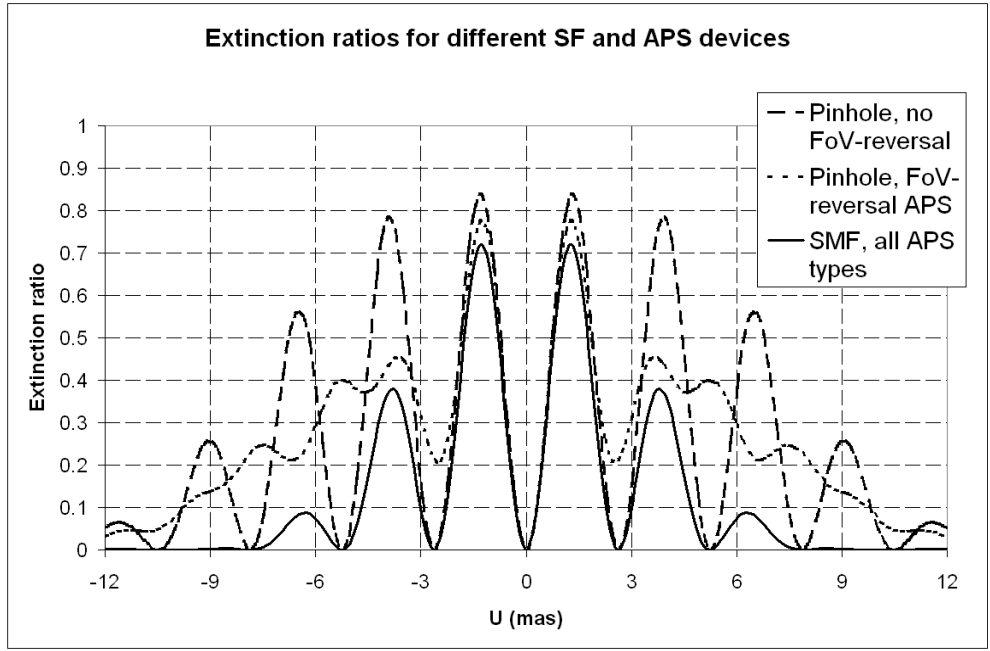

Fig. 5. Slices along U-axis of extinction maps for different SF and APS devices (dashed line: pinhole filtering with no FoV reversal; dotted line: pinhole filtering with FoV reversal; solid line: SMF filtering)

## 4. Conclusion

In this paper was discussed the influence of some subsystems on the potential performance of a nulling interferometer, in terms of its extinction map projected on-sky. Of particular importance are the spatial filtering and achromatic phase shifting devices, whose basic concepts may strongly affect the rejection ratios achieved within the interferometer FoV. We first derived a set of equations involving cross correlation products and allowing to compute digitally the nulling maps generated by a dual-telescopes Bracewell interferometer. We believe that the obtained relationships, appearing as the product of an on-sky fringe pattern with an envelope function depending on the SF and APS designs, can be generalized to other multi-aperture interferometers, at least if no FoV-reversal APS are used. Numerical simulation were carried out, showing significant differences between the calculated extinction maps – also leading to alleviate some conclusions in Ref. [16]. As an empiric rule, spatial filtering with pinholes seems to produce brighter and more numerous constructive fringes than single-mode optical fibers do, consequently achieving larger nulling maps and actual interferometer FoV. It must be highlighted, however, that only basic hypotheses were considered herein (monochromatic light, Bracewell configuration with two apertures), thus the conclusion may somewhat differ when broad spectral bands are considered on multiple apertures. SMF, in particular, are reputed for being less sensitive to chromaticity than pinholes, thus additional sets of numerical simulations involving full operational spectral widths of nulling interferometers are required. But if our present results are confirmed, pinhole filtering would be more efficient for serendipitous planet discoveries, while the use of single-mode fibers would be limited to the characterisation of already known extra-solar planets or systems. In any case, it has to be underlined that simulations of full-field extinction maps are of prime importance when designing a nulling interferometer, whose performance cannot be completely characterized by means of its sole maximal attainable nulling rate.

**Acknowledgement**

The author would like to thank the following people for nice technical discussion, e-mail exchanges and helpful remarks: Marc Barillot, Frédéric Cassaing, Charles Hanot, Dimitri Mawet, Denis Mourard, Yves Rabbia and the anonymous reviewer.